\documentstyle[12pt,psfig]{article}

\begin{document}
\newcommand{\ECM}{$^1$\em Departament d'Estructura i Constituents de la
Mat\`eria, Facultat de F\'\i sica\\
Universitat de Barcelona, Diagonal 647, E-08028 Barcelona, Spain \\
                       $\phantom{and}$                        \\
$^2$ Department of Physics, University of Oslo, P.O. Box 1048 Blindern,
N-0316 Oslo, Norway }

\def\thefootnote{\fnsymbol{footnote}}
\pagestyle{empty}
{\hfill \parbox{6cm}{\begin{center} hep-th/9711xxx\\
                                    November 1997
                     \end{center}}}
\vspace{1.5cm}

\begin{center}
\large{On RG potentials in Yang-Mills Theories}

\vskip .6truein
\centerline {J.I. Latorre$^1$\footnote{e-mail: latorre@ecm.ub.es}
and C.A. L\"utken$^2$\footnote{e-mail: c.a.lutken@fys.uio.no}}
\end{center}
\vspace{.3cm}
\begin{center}
\ECM
\end{center}
\vspace{1.5cm}

\centerline{\bf Abstract}
\medskip
We construct an RG potential for $N=2$ supersymmetric $SU(2)$ Yang-Mills
theory, and extract a positive definite metric by comparing its gradient
with the recently discovered beta-function for this system, thus
proving that the RG flow is gradient in this four-dimensional field theory.
We also discuss how this flow might change after supersymmetry breaking,
provided the quantum symmetry group does not, emphasizing the non-trivial
problem of asymptotic matching of  automorphic functions to perturbation
theory.  

\newpage
\pagestyle{plain}

Our understanding of the non-perturbative aspects of non-abelian gauge theories
is at this point in time very limited.  Most of our knowledge of
the strongly coupled domain comes from exploiting approximate global continuous
symmetries, whose origin are not well understood and may be coincidental,
to construct low-energy effective field theories encoding the dynamics of
the effectively massless modes whose existence is guaranteed by Goldstone's
theorem.
Recently it has been realized that certain theories may possess exact global
discrete quantum symmetries which are so constraining that the theory can
essentially be solved exactly, even in the non-perturbative domain.
Such symmetries may underlie the remarkable phase- and fixed point structure
of the subtle quantum Hall system \cite{LR},
and their discovery \cite{SW} in $N=2$ supersymmetric
YM theories has led to a deep understanding of these systems, recently
culminating in the construction of an exact representation of the contravariant
(physical) beta-function \cite{beta}.
It is the purpose of this letter to further explore the consequences of such
symmetries and their relationship to supersymmetry in non-abelian gauge
theories.

The main idea to be discussed here, originally devised for the quantum Hall
system \cite{BL},  is that the combined constraints of a quantum symmetry
group (contained in the modular group)
and matching to perturbation theory in the asymptotically flat domain
(weak coupling) may be sufficient to completely harness the beta-function.
In particular, automorphy of the contravariant beta-function is incompatible
with holomorphy on the entire fundamental domain, and the way in which
the analytical structure must be relaxed is indicated by the asymptotic
behaviour of the beta-function:
\begin{equation}
\beta^z = \frac{dz}{dt} =
b_0 + \frac{b_1}{y} + \frac{b_2}{y^2} + \dots + (q{\rm-expansion}),
\label{betafn}
\end{equation}
where $t$ is a scale parameter and for YM theory
$z = x + iy = \theta/2\pi + 4\pi i/g^2$, and $q(z) = \exp(2\pi iz)$.

If the theory is $N=2$ supersymmetric, a strong non-renormalization theorem
implies that there exists a renormalization scheme in which only one loop
contributes, so that all $b_i$ except $b_0$ vanish \cite{Seiberg}.  Hence the
beta-function is asymptotically holomorphic, and analyticity
is only violated by a simple pole  at $z=0$, which is the unique point
on the boundary of the fundamental domain where the effective field theory
breaks down.   This is as close to
holomorphic as a (sub-)modular beta-function can get, and its form is then
completely fixed by the discrete quantum symmetry and asymptotic matching
at $z\rightarrow i\infty$ (to perturbation theory and a single-instanton
calculation).

This idea\footnote{Similar ideas have been discussed in Ref.\cite{Haagensen}
in the context of non-linear sigma models.}
was very recently \cite{Ritz} applied to the contravariant
beta-function for $SU(2)$ and $N=2$ invariant YM theory \cite{beta},
thus simplifying its derivation enormously.
We first extend this line of reasoning to the $N=2$ covariant
beta-function, which by comparison with the contravariant one allows us
to extract a positive definite metric with reasonable physical properties.
Since the covariant beta-function is gradient, this proves that the
RG flow in this system is gradient \cite{cthm}.

If the theory is not supersymmetric,  $b_1$ does not vanish in any scheme
and we see that the analytic structure changes dramatically.
It seems unlikely that such a function could still be automorphic,
but a remarkable ``quasi-holomorphic'' exception exists, and this
can be used to build a
candidate beta-function with similar uniqueness properties to the
supersymmetric case.  Thus, it is not inconsistent to discuss the
properties of non-supersymmetric beta-functions automorphic under
quantum symmetries.  If such symmetries do exist they may or
may not be survivors of supersymmetry breaking,
but it seems plausible that it is the interplay of
supersymmetry, quantum duality and holomorphy which must be elucidated if
the spectacular success of $N=2$ theories is to be extended to
non-supersymmetric models.

\bigskip

A simple yet precise picture of the RG properties of $N=2$ $SU(2)$ YM has
recently emerged from the explicit construction of an exact non-perturbative
beta-function \cite{beta}:
\begin{equation}
\label{betafunction}
\beta^z(z) = \frac{1-4f(z)}{f'(z)} =
-{i\over \pi}
\left( \frac{1}{\theta_3(z)^4} +\frac{1}{\theta_4(z)^4}\right)
\label{betaN=2}
\end{equation}
where $f = -(\theta_3\theta_4/\theta_2^2)^4$ is invariant under the quantum
symmetry group $\Gamma_T(2)$ of this system, and $\theta_i (i=2,3,4)$
are the conventional elliptic theta-functions, related by
$\theta_2^4 = \theta_3^4 - \theta_4^4$ \cite{Rankin}.  This
infinite non-abelian discrete quantum symmetry group is the largest sub-group
of the modular group which contains translations by one. This is why it is
$\Gamma_T(2)$ which appears both in YM and in the quantum Hall system.
Translations by one\footnote{We follow the conventions of Rankin
\cite{Rankin} as closely as possible, but adopt the standard physics notation
for the generators of the modular group $SL(2,Z)$. Thus, translations are
denoted by $T$ (rather than $U$), and simple duality $z\rightarrow -1/z$ is
denoted by $S$ (rather than $V$).}
$T: z\rightarrow z+1$ is a symmetry because the theta parameter is periodic
under shifts by $2\pi$, and guarantees that all functions automorphic under
this group have $q$-expansions with $q=\exp(2\pi iz)$.
The other generator of the group can be conveniently chosen as $ST^2S$.

Very recently a simple and illuminating derivation of this beta-function was
given by Ritz \cite{Ritz}, which relies on the observation that
the contravariant beta-function transforms like a weight $w = -2$ automorphic
function\footnote{Mathematical textbooks like to reserve the
words ``function'' and ``form'' for automorphic objects with special
analyticity properties, but their usage is inconsistent and confusing, so
we do not qualify our language in this way.  We shall therefore have to be
explicit about the number of poles possessed by our functions.}:
\begin{equation}
\beta^z(\gamma(z)) = \frac{d\gamma}{dz} \beta^z(z) = (cz + d)^{-2}\beta^z(z)
\label{betatransf}
\end{equation}
for any $\gamma\in \Gamma_T(2)$ acting holomorphically by fractional linear
transformations on the modular parameter $z$:  $\gamma(z) = (az+b)/(cz+d)$.
As observed in \cite{BL}, together with asymptotic matching to the
perturbatively evaluated beta-function, this is a powerful constraint on the
possible forms of this function.  In the $N=2$ case, where the beta-function is
meromorphic, it does in fact uniquely fix the beta-function to the one
given above, up to a single constant which will be discussed shortly.

We now adapt the argument to the covariant beta-function, which transforms
as a weight $w = +2$ function.  The first observation
(Rankin \cite{Rankin}, p.111) is that for groups of
genus zero (which includes $\Gamma(1) = SL(2,Z)$ and $\Gamma_T(2)$) there
exists an invariant ($w = 0$) meromorphic function with a single pole
(and a single zero) in the fundamental domain, which we have called $f$ above.
Any other weight zero function with total order $q$ of zeros (or poles) is
given by a rational function $P(f)/Q(f)$, where the degree in $f$ of both the
polynomials $P$ and $Q$ do not exceed $q$.  Applying this theorem to the
invariant function $g = \beta^z f'$,  and using the
asymptotic behaviour of $f'$ and the $N=2$ beta-function,
it follows that $\beta^z = (c_1 +c_2 f)/f'$, where
the conventional choice of scheme corresponds to $c_2=-4$
(from perturbation theory) and $c_1=1$ (one-instanton computation).

Similarly, assuming that there exists an asymptotically flat metric,
the same argument applied to the invariant function $h = \beta_z/f'$ gives
\begin{equation}
\beta_z = {f'\over 1 - 4f} =\partial_z\Phi \qquad
{\rm with}\qquad \Phi = -{1\over 4} \log\left| 1-4f\right|^2\ .
\label{cfunction}
\end{equation}
A contour plot of the RG potential is given in Fig. 1, where the horizontal
axis is $x = {\rm Re} z = \theta/2\pi$ and the vertical axis is
$y = {\rm Im} z = 4\pi/g^2$. 

\begin{figure}[htb]
\begin{center}
\mbox{\psfig{figure=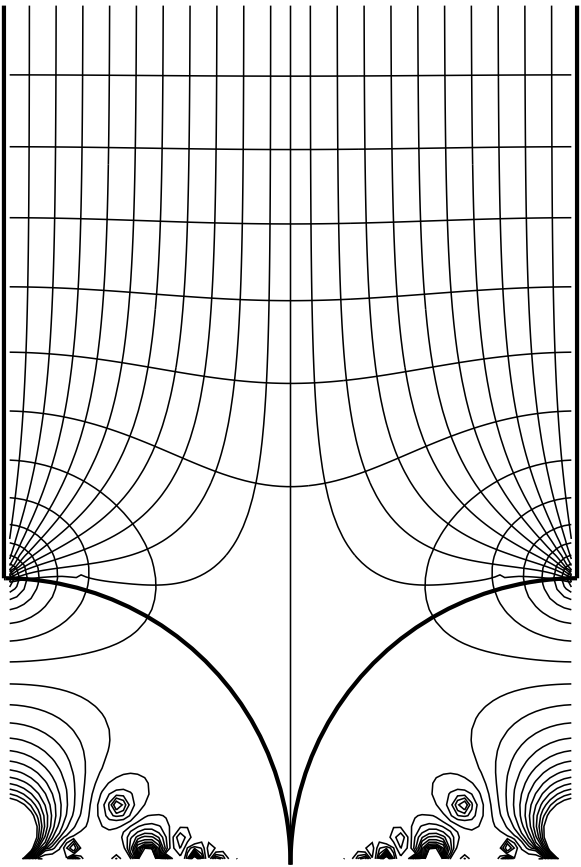,width=8cm,angle=0,height=12cm}}
\end{center}
\caption{\footnotesize {The phase and RG flow diagram of $N=2$ supersymmetric
$SU(2)$ Yang-Mills theory, obtained from the RG potential $\Phi$ exhibited in
Eq.(\ref{cfunction}).  Flow lines are downward and displayed only in the 
fundamental domain, while equipotential lines are shown for $\theta$ 
between $-\pi$ and $\pi$.}}
\label{fig1}
\end{figure}
A similar flow was very recently displayed in Ref.\cite{Dolan}.
It is remarkable that this RG potential for $N=2$ YM is so simply related
to the unique univalent holomorphic function $f$ in $\Gamma_T(2)$. From
this construction we can further extract a metric
\begin{equation}
G_{z\bar z} = \beta_z \beta_{\bar z} = \left|\frac{f'}{1-4f}\right|^2
= \partial_z \partial_{\bar z} K\qquad {\rm with}\qquad
K = c \Phi + \frac{1}{2}\Phi^2
\label{metric}
\end{equation}
which behaves correctly under the action of the Lie derivative \cite{BL},
and is flat, finite and positive definite except at the fixed points.
The K\"ahler potential $K$ can have any linear term since 
$\partial_z\beta_{\bar z} = 0$.
At $z=i\infty$ it is finite, as behooves a metric adapted to a basis of
noninteracting operators in the perturbative domain. Furthermore, it
only carries mixed components, as expected from the OPE of YM in
the perturbative domain, since $\langle(FF)(F^*F)\rangle = 0$.
At the IR fixed point,
 $z = (1 + i)/2$, it has a coordinate singularity, which is to be expected
if it is adapted to the perturbative basis of weakly interacting operators.
At strong coupling we expect a different (``dual'') basis of weakly
interacting composite operators to be more convenient, so it is natural for
the metric adapted to the perturbative domain to have a coordinate singularity
in the infrared.
The curvature is zero, except at $z=0$, where the metric is ill defined
(not positive).  This is also to be expected since this is where the
effective field theory breaks down due to the presence of new massless modes.
The Seiberg-Witten metric is quite different.  It also has an essential
singularity at $z=0$, but it has a coordinate singularity in the perturbative
domain, and is well behaved at the IR fixed point, indicating that  it is
better adapted to describing strong coupling.

These issues are not of any great relevance for us, since we have now
achieved our goal:  by exhibiting a potential function and a positive
definite metric, we have shown that the RG flow is gradient.
More precisely, we have just specified the elements in the following chain
of equations:
\begin{equation}
-\beta^i \partial_i \Phi=-\beta^i\beta_i=-\beta^i\beta^j G_{ij}\le 0\ ,
\end{equation}
where the inequality is obtained because $G_{ij}$ is positive definite.
It is worth noting that this is the first explicit construction of a
non-perturbative function of this kind for a  $d=4$ field theory.
The much harder question of whether or not these functions
have physical interpretations as the C-function, which
counts effectively massless degrees of freedom at fixed points, and the
Zamolodchikov metric, which is given by the correlators of relevant and
marginal operators, is logically independent. All efforts addressing the
irreversibility of RG flows are ultimately motivated by
such physical considerations.  Different approaches
\cite{cthm4d} have recently converged on a solid
candidate for the c-number (central charge) of the conformal field theories
appearing at non-trivial fixed points.  It is the coefficient
of the Euler density in the trace anomaly in curved space
(labeled as $a$ in \cite{ Dan} and $\beta_b$ in \cite{ OL}), and
this candidate has been shown, empirically, to be positive and to
decrease along RG flows in a large number of non-trivial examples
\cite{Bastianelli,Dan}.  Further work has related the sign of $\beta_b$ to
a new form of the weak energy positivity theorem \cite{OL}.

Although we at present are unable to relate $\Phi$ to a distinct
physical observable, this is not required in order to prove that the RG flow
is gradient. We are also not able to relate our metric $G_{z\bar z}$
to the OPE of the operators in this theory.  So, while it has many good
properties, being non-singular and automorphic as well as having correct
asymptotics and Lie transport, we do not know how to relate it to the basic
correlators of the theory.

\bigskip

Let us now turn to the issue of scheme dependence \cite{finnell}
alluded to above.
As we have seen,  the $\Gamma_T(2)$-symmetry reduces the
mathematical freedom in the construction of the $\beta^z$-function
to finding the appropriate values for $c_1$ and $c_2$. It was
argued in Ref.\cite{Ritz} that $c_2$ controls the asymptotic behaviour
of $\beta^z$, and therefore is fixed by the (one-loop) perturbative
coefficient $b_0 = -2i/\pi$ of the beta-function, while
$c_1$ was fixed by a one-instanton computation.  It is obvious that
$c_2$ is scheme independent. Let us argue that $c_1$ does not bring any
scheme dependence either. A change of scheme corresponds to a finite
redefinition of coupling constants which, in turn, leads to a
different value of the RG  invariant:
\begin{equation}
\Lambda=\mu e^{-\int^{g(\mu)} \beta^{-1}}.
\end{equation}
In our case we find:
\begin{equation}
\left({\Lambda\over\mu}\right)^{-c_2} = c_1 + c_2 f
= {-c_2\over 256 q^2}+ \left( c_1 +{3 c_2 \over 32 }\right)
 - {69 c_2\over 64}  q^2 + \dots .
\end{equation}
Although $c_1$ appears as a constant on the rhs of the above equation
and looks suspiciously similar to the lhs, it cannot be
absorbed by a finite redefinition of the coupling $z$. Furthermore,
by plotting $\Phi = c_2^{-1}\log \left| c_1 + c_2 f\right|^2$ for
different values of $c_1$ it is immediately evident that this function
yields flows with different topologies, moving the IR attractor all
over the $z$-plane. Therefore, rather than parametrizing different schemes,
$c_1$ labels different universality classes: $c_1 = 1$ corresponds to $N=2$
YM (see Fig. 1),  $c_1 = 0$ leads to flows of the type discussed below
(see Fig. 2), while other values of $c_1$ give exotic flows with
no obvious interpretation.

In short, the contravariant beta-function for $N=2$ $SU(2)$ YM is completely
fixed by asymptotic matching to a one-loop and a one-instanton computation.
The scheme is selected in such a way that
the beta-function takes its simplest form with respect to modular
transformations.  Coupling constant reparametrizations will
produce a change of scheme where the entanglement of the perturbative
and non-perturbative parts of the beta-function can be very complicated.
A deeper understanding of scheme
dependence and the way in which reparametrizations of the moduli space
can be used to simplify the description of a C-theorem is needed.

\bigskip

Finally we consider the possibility that non-supersymmetric gauge theories are
also constrained by a quantum duality.  The simplest origin of such a quantum
symmetry $\Gamma$ would be if the $N=0$ theory is a broken version of
$N=2$, and $\Gamma = \Gamma_T(2)$ is sufficiently robust to survive
supersymmetry breaking
(see the relevant instance proposed
 in \cite{luis}), but we do not need to commit ourselves to this
scenario.  There are several curious properties of ordinary YM gauge theory
that may be construed as hinting at such a structure.  The natural
parametrization of YM moduli space is obtained by writing the action
in terms of self-dual $(F_+)$ and anti-self-dual $(F_-)$ field
configurations: $L = z F_+^2 + \bar z F_-^2$, where
$z$ is the same complexified coupling constant that  appears in the 
$N=2$ case.  Instanton corrections to the beta-function
have been argued \cite{russians} to be  periodic in theta and
exponentially suppressed in perturbation theory, in precisely such a way
that they are holomorphic in this parameter:
$q = \exp(i\theta - 8\pi^2/g^2) = \exp(2\pi iz)$, as required by the
$q$-expansion of automorphic functions.  Note also that the $\theta$-parameter
is ill-defined at the fixed point $z=i\infty$ of the (sub-)modular group,
as required by perturbation theory.
Furthermore, we recall that the original introduction of the modular
group into physics was precisely in an attempt to understand the
oblique confinement scenario advocated by t'Hooft\cite{Cardy}.
Unless this remarkable confluence of convenient results is mere coincidence,
it is not unnatural to conjecture that they are orchestrated by a
hidden quantum symmetry contained in the modular group.
Since the vacuum is invariant under translations of the theta-parameter
by $2\pi$, $T$ is one of the generators and it follows that if
$\Gamma$ is not too small then $\Gamma = \Gamma_T(2)$.

\begin{figure}[htb]
\begin{center}
\mbox{\psfig{figure=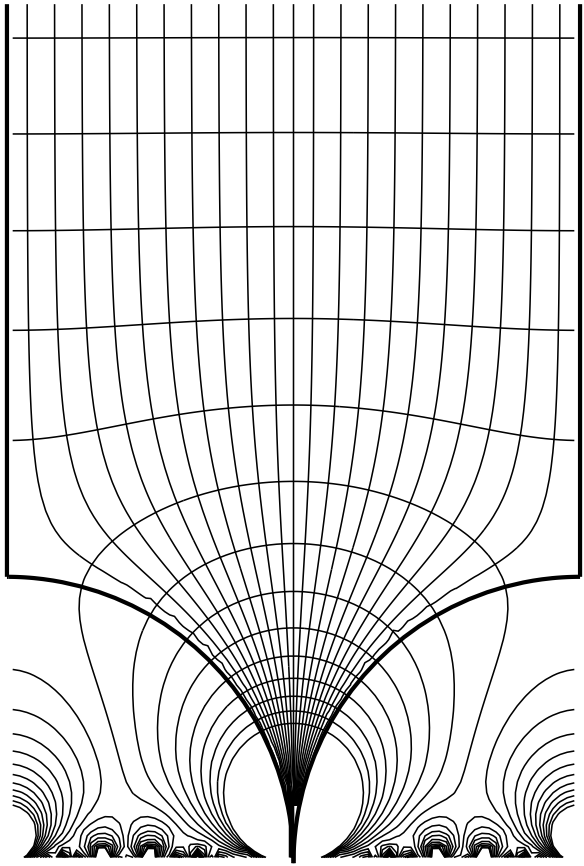,width=8cm,angle=0,height=12cm}}
\end{center}
\caption{\footnotesize {The phase and RG flow diagram of the potential
$\Phi_E$.  Flow lines are downward and displayed only in the fundamental
domain, while equipotential lines are shown for $\theta$ between $-\pi$
and $\pi$.}}
\label{fig2}
\end{figure}

We now formulate a precise conjecture which implies a unique form for the
RG potential, and therefore the phase- and RG flow diagram in the
non-supersymmetric case.  We exploit the remarkable mathematical fact that
for $\Gamma_T(2)$ there exist two basic functions that transform properly with
weight $w = +2$ \cite{Rankin,BL}:
\begin{eqnarray}
E_2^T(z) &=& \partial \Phi_E(z,\bar z) = 1 + (q{\rm-expansion})\nonumber\\
H_2^T(z,\bar z) &=& \partial \Phi_H(z,\bar z) = 
\frac{1}{\pi y} + (q{\rm-expansion}) \nonumber\\
\end{eqnarray}\nonumber
where, curiously, the potentials are built only from the lowest weight
cusp forms (``instanton forms'') for $\Gamma_T(2)$ and $\Gamma(1)=SL(2,Z)$,
and $y = {\rm Im} z$.  They can be conveniently written as:
\begin{equation}
\Phi_E(z,\bar z) = \frac{i}{\pi} \log\vert f\vert, \qquad
\Phi_H(z,\bar z) = \frac{i}{2\pi}\log\vert y^4\theta_3^8\theta_4^8 \vert .
\end{equation}
Notice that while both these functions are fully automorphic, and
$E_2^T(z)$ is holomorphic, $H_2^T(z,\bar z)$ is {\it not}.  It is the
unlikely existence of the latter, which we call a ``quasi-holomorphic''
or ``Hecke'' function, which allows us to formulate our conjecture.
A contourplot of $\Phi_E$ is shown in Fig. 2, and the plot for
$\Phi_H$ is very similar.

Our conjecture for the non-supersymmetric case can now be stated succinctly.
If the quantum symmetry group is $\Gamma_T(2)$,
and the holomorphic structure is violated only in the
benign way exhibited above in the beta-function and in the Hecke form,
by admitting only non-analytical terms of the type $1/y$ in the
asymptotic domain (`t Hooft scheme, where only $b_0$ and $b_1$ are 
different from zero), then the candidate covariant beta-function is 
a linear combination of $E_2^T$ and
$H_2^T$ where one of the constants can be fixed by comparison with
perturbation theory, provided the contravariant beta-function is constructed
using an asymptotically flat metric.   Thus, the final form of the
conjecture is that $\Phi = b_0 \Phi_E + b\Phi_H$, where $b$ conspires
with the sub-leading term in the metric to give $b_1$ etc. The phase- 
and RG flow diagram implied by this potential is evident from Fig. 2,
showing an IR fixed point at $z=0$, except when $\theta = \pi$ in which case
$\theta$ is not renormalized and the IR fixed point is a critical point at
$\alpha_s = g^2/4\pi = 2$. So this potential has the desirable property of
having only one phase in which the theta-parameter renormalizes
to zero in the IR domain, unless theta is exactly $\pi$.  This suggests 
that the strong CP problem may be automatically resolved 
within YM theory by quantum duality, 
since it forces theta to flow towards zero at strong coupling, 
thus eliminating the need for axions or other ad-hoc constructions. 
The running of the theta-parameter would stop
at the mass gap, yielding a finite but very small value of the physical
theta-parameter.  The existence of the critical point is in our context
inescapable, and it may be possible to test this if the region with theta near
$\pi$ can be probed in lattice simulations of $SU(2)$ Yang-Mills.

\bigskip
\centerline{\bf Acknowledgments}

We are grateful to J. Ambjorn, C. Burgess, D. Z. Freedman,
P. E. Haagensen and F. Quevedo for useful discussions and 
constructive comments on the manuscript.
Financial support from CICYT, contract AEN95-0590,
and from CIRIT, contract GRQ93-1047, and the Norwegian Science Council
are acknowledged.


\begin{thebibliography}{99}
\bibitem{LR} C.A. L\"utken and G.G. Ross, {\sl Phys. Rev.} {\bf B45}
(1992) 11837; {\bf B48} (1993) 2500; C. A. L\"utken, {\sl Nucl. Phys.}
{\bf B396} (1993) 670

\bibitem{SW} N. Seiberg and E. Witten, {\sl Nucl. Phys.} {\bf B426}
(1994) 19

\bibitem{beta} J.A. Minahan and D. Nemechansky, {Nucl. Phys.} {\bf
B468} (1996) 72; G. Bonelli and M. Matone, {\sl Phys. Rev. Lett.} {\bf
76} (1996) 4107

\bibitem{BL} C.P. Burgess and C.A. L\"utken, {\sl Nucl. Phys.} {\bf B500}
(1997) 367

\bibitem{Seiberg} P. S. Howe, K.S. Stelle and P. West, {\sl Phys. Lett.}
{\bf B123} (1983) 55;  N. Seiberg, {\sl Phys. Lett.} {\bf B206} (1988) 75

\bibitem{Haagensen} P.E. Haagensen, {\sl Phys. Lett.} {\bf B382}
(1996) 356; K. Olsen, {\sl Nucl. Phys.} {\bf B504} (1997)326;
K. Olsen and R. Schiappa, {\sl Phys. Rev. Lett.} {\bf 79} (1997) 3573;
P.E. Haagensen and K. Olsen, {\sl hep-th/9704157};
P.E. Haagensen, K. Olsen and R. Schiappa, {\sl hep-th/9705105}

\bibitem{Ritz} A. Ritz,  {\sl hep-th/9710112}

\bibitem{cthm} A.B. Zamolochikov, {\sl JETP Lett.} {\bf 43} (1986) 730

\bibitem{Rankin} R.R. Rankin, {\it Modular Forms and Functions}
(Cambridge University Press, 1977)

\bibitem{Dolan}  B.P. Dolan, {\sl hep-th/9710161} 

\bibitem{cthm4d}J.L. Cardy, {\sl Phys. Lett.} {\bf B215} (1988) 749;
H. Osborn, {\sl Phys. Lett.} {\bf B222} (1989) 97;
A. Cappelli, D. Friedan, J.I. Latorre, {\sl Nucl. Phys.} {\bf B352}
(1991) 616

\bibitem{Dan} D. Anselmi et al., {\sl hep-th/9708042};
D. Anselmi et al. {\sl hep-th/9711035}

\bibitem{OL} H. Osborn and J.I. Latorre, {\sl  hep-th/9703196}, to
appear in {\sl Nucl. Phys. B}

\bibitem{Bastianelli} F. Bastianelli, {\sl Phys. Lett.} {\bf B369} (1996) 249

\bibitem{finnell} D. Finnell and P. Pouliot, {\sl Nucl. Phys.} {\bf B453}
(1995) 225

\bibitem{luis} L. Alvarez-Gaum\'e et al., Int. J. Mod. Phys. A11 (1996)
4745

\bibitem{russians} V.G. Knizhnik and A. Yu. Morozov, {\sl JETP Lett.}
{\bf 39} (1984) 240

\bibitem{Cardy} G. t'Hooft, {\sl Nucl. Phys.} {\bf B190} (1981) 455;
J.L. Cardy and E. Rabinovici, {\sl Nucl. Phys.} {\bf B205} (1982) 1; 
J.L. Cardy, {\sl Nucl. Phys.} {\bf B205} (1982) 17

\end{thebibliography}
\end{document}